\documentclass[doublecol]{epl2}

\newcommand{\cU}{{\cal U}}
\newcommand{\cG}{{\cal G}}

\usepackage{epsfig}
\usepackage{graphicx}
\usepackage{amsmath}
\usepackage{amssymb}
\usepackage{dcolumn}
\usepackage{amsmath}
\usepackage{amsfonts}

\title{Dynamical localization of gap-solitons by time
periodic forces} \shorttitle{Dynamical localization of
gap-solitons}
\author{Yu. V. Bludov\inst{1} \and V. V. Konotop\inst{2,3}
\and M. Salerno\inst{4}}\shortauthor{Yu. V. Bludov, V. V. Konotop, M. Salerno}

\institute{ \inst{1}Centro de F\'{\i}sica,
Universidade do Minho, Campus de Gualtar, Braga 4710-057, Portugal
\\
\inst{2}Centro de F\'{\i}sica Te\'orica e Computacional,
Universidade de Lisboa, Complexo Interdisciplinar, Avenida
Professor Gama Pinto 2, Lisboa 1649-003, Portugal
\\
\inst{3}Departamento de F\'{\i}sica, Faculdade de Ciências,
Universidade de Lisboa, Campo Grande, Edifício C8, Piso 6, Lisboa
1749-016, Portugal
\\
\inst{4}Dipartimento di Fisica "E. R. Caianiello," and
Consorzio Nazionale Interuniversitario per le Scienze Fisiche
della Materia (CNISM), Universit\'a di Salerno, I-84081, Baronissi
(SA), Italy}

\abstract{ The phenomenon of dynamical
localization of matter wave solitons in optical lattices is first
demonstrated and the conditions for its existence are discussed. In
addition to the trapping linear periodic potential we use a
periodic modulation of the nonlinearity in space to eliminate
nonexistence regions of gap-solitons in reciprocal space. We show
that when this  condition is achieved, the observation of
dynamical localization in true nonlinear regime becomes possible.
The results apply to all systems described by the periodic
nonlinear Schr\"odinger equation, including Bose-Einstein
condensates of ultracold atoms trapped in optical lattices and
arrays of waveguides  or photonic crystals in nonlinear optics.}

\begin{document}
\maketitle An interesting phenomenon occurring in periodic
nonlinear systems is the possibility to sustain  different types
of localized waves  for arbitrarily long times due to the interplay
between periodicity, dispersion and nonlinearity.
 In particular, optical lattices (OL) can support
so called gap-solitons, i.e. localized solutions of the periodic
nonlinear system with energies
located inside forbidden gaps of the spectrum. These solutions
exist both for focusing (self-attractive) and defocussing
(self-repulsing) interactions due to the coupling induced by the
nonlinearity between forward and backward propagating waves
undergoing Bragg scattering by the periodic structure
\cite{mills}. Gap solitons can be found in many physical systems
including photonic crystals and layered structures \cite{sterke},
arrays of optical fibers \cite{Mandelik} and Bose-Einstein
condensates (BECs) in OL. In this last case, gap-solitons have
been experimentally observed in \cite{Ober} for BEC with positive
scattering lengths (repulsive interactions) and their properties
have been characterized both for discrete (BEC arrays)
\cite{TS01,ABDKS01} and for continuous BEC \cite{KS02,AKS02} (see also reviews
\cite{BK,reviews,PhysD}). As stationary states with energies
inside band gaps they are intrinsically localized and
remain stable for long time. By applying  external forces, such as
the one generated by the gravity (for BEC in periodic vertical
traps) or by the acceleration of the OL, solitons can be forced to
move. A moving soliton however has a frequency
belonging to an allowed band. It is not any more an exact
solutions of the one-dimensional Gross-Pitaevskii equation with a
periodic potential, but can be viewed as a persisting wave
packet.

 On the other hand,  it is known from the linear
theory, that a time periodic force  applied to a quantum  particle
in a lattice can result in the phenomenon of dynamical
localization~\cite{Dunlap}. In the simplest terms this effect can
be viewed as a resonance which occurs when the strength of the force and its frequency
take certain discrete values. Then a particle undergoes
oscillatory motion in a finite spatial domain (namely in this
sense  the localization is understood).  Outside of these resonances
the particle acquires a drift velocity which enables transport.
For sinusoidal time dependent forces resonances are found to be
roots of the zero-order Bessel function. On a pure quantum mechanical level~\cite{Holthaus} dynamical localization can also be characterized as a complete  suppression of the inter-well tunneling\footnote{The complete suppression of inter-well tunneling is a valid statement only in the high-frequency limit}  of the
particle in a periodic potential~\cite{EWH05,Morsch}.

The linear dynamical localization has been
experimentally observed
in several systems including arrays of curved optical
waveguides~\cite{optics} and BEC  trapped~\cite{Morsch,Eckardt1}.
Properties of the ac forces (electrical fields) that result in the
dynamical localization of an electron in a periodic potential were
theoretically investigated in \cite{dignam} where it was shown
that for generic (i.e. beyond tight-binding) band structures exact
dynamical localization is possible only for electric field
displaying discontinuities at all changes of the sign of the field
(the dynamical localization phenomenon, however,  was shown to be
surprisingly tolerant to the smoothing of these discontinuities
\cite{Domachuk}).

Thus, while the concept of a gap-soliton is
intrinsically nonlinear the concept of dynamical localization is
independent on nonlinearity. At the same time, the arguments
applicable to a single particle do  not mean of course that a wave
packet composed of oscillating atoms will oscillate without
dispersion, i.e. preserving its width. Then the natural question
arises: {\it Is it possible to observe  dynamical localization of
a non spreading wave packets, i.e. of gap-solitons}? Possibility
of positive answer to this question stems from the results
reported in Ref \cite{SKB08}, where it  was shown that properly
designed nonlinearity can guarantee long-lived Bloch oscillations
of gap-solitons of small amplitude originated by constant forces.

 The aim  of the present Letter is to show for the
first time how the dynamical localization of gap-solitons of the
continuous periodic nonlinear Schr\"odinger (NLS) models can be
achieved  by means of suitable spatial modulations of the
nonlinearity (nonlinear OL). For BEC in OL this modulation can be
obtained  by changing the scattering length along the condensate
using the optically induced Feshbach resonance technique. We shall
use the spatial modulation to suppress (or strongly reduce) non
existence regions for gap-solitons in the BZ. As a result we have
that the resonances for the dynamical localization of gap-solitons
are the same as the one obtained in the seemingly very different
linear system.

 An intuitive understanding of why the oscillatory
motion of single particles is accompanied by the spreading out of
the wave packets and why this situation can be inverted in
presence of space dependent nonlinearities can be obtained from
the analysis of nonlinear discrete models. It is known that the
usual (e.g. physical) discretization of the NLS equation (DNLS)
\cite{AKKS}
\begin{equation}
i\dot{c}_{n}=\omega_{0}c_n+\omega_{1}(c_{n+1}+c_{n-1})+\chi
|c_n|^2 c_{n} \label{eq:discrete}
\end{equation}
has the onsite nonlinearity and  does not display any dynamical
localization phenomenon in presence of  ac linear forces (any
initially localized wave packet quickly spread out under the
action of the force). In contrast with its continuous limit, Eq.
(\ref{eq:discrete}) is nonintegrable,  having only two conserved
quantities: the energy and the norm. In BEC contexts this equation
arises as a tight binding model  of the NLS equation and is used
to describe arrays of coupled BECs \cite{TS01,ABDKS01}. The
integrable discrete version of the NLS equation, known as the
Ablowitz-Ladik (AL) model, has the nonlinearity non local i.e.
splitted on two adjacent sites: $\chi |c_n|^2(c_{n-1}+ c_{n+1})$,
(which must be substituted in (\ref{eq:discrete}) instead of
$|c_n|^2 c_{n}$) and it is known that it does display the
dynamical localization phenomenon. The first observation of
dynamical localization for discrete solitons was indeed achieved
for this model \cite{KCV,Mario}. The generalization of the theory
of dynamical localization by including the one-site nonlinearity,
as well the understanding of the physical picture, was suggested
in~\cite{Mario} using the discrete nonlinear Schr\"odinger (DNLS)
equation as a governing model, where again the same frequencies
leading to the dynamical localization were obtained.
\begin{figure}
\centerline{
\includegraphics{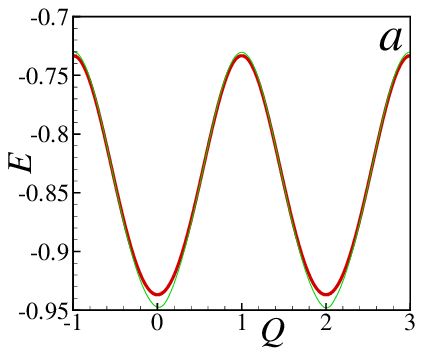}
\includegraphics[width=4.cm,height=3.6cm,angle=0,clip]{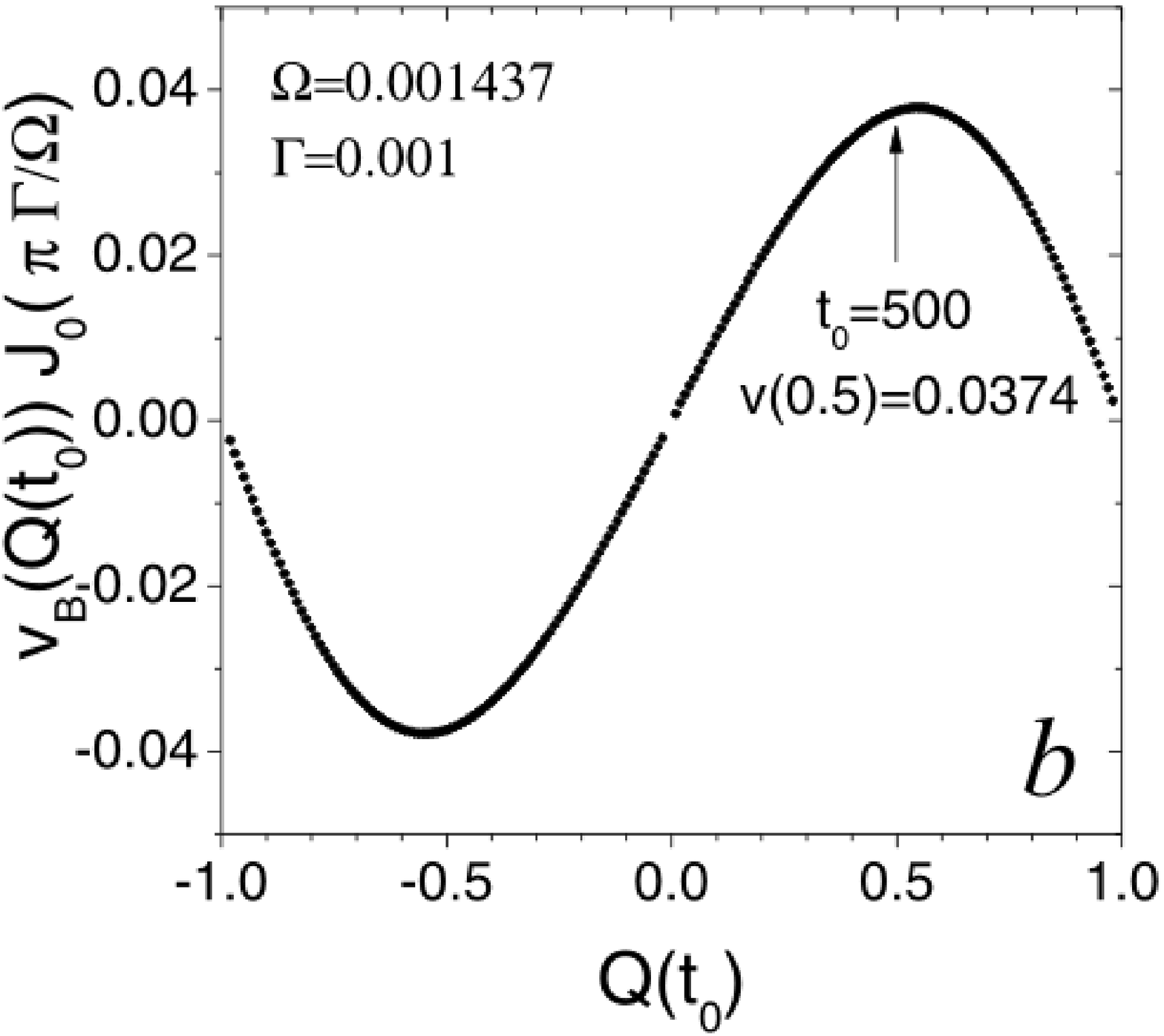}}
\caption{(a) The lowest linear band energy $E$ \textit{vs} crystal
momentum $Q$ (bold line) and the evolution of energy of the
soliton with norm $N=5.42$ during its movement under the constant
external force $\gamma(t)=\Gamma=0.001$ (thin line). The evolution
of the soliton crystal momentum is $Q(t)=-1+\Gamma t$; (b) The
drift velocity as a function of the initial crystal momentum of
the gap-soliton  for the nonresonant frequency $\Omega=0.001437$.
The linear and nonlinear OLs are fixed as $\cU(x)=-3\cos(2x)$ and
$\cG(x)=-0.777+\cos(2x)$.} \label{fig:ms0}
\end{figure}

To understand why the dynamical localization phenomenon
of undistorted spatially localized wave packets
occurs in the AL model and does not occur in the DNLS chain it is of
interest to consider how the stability  of linear modes of the
underlying discrete linear chains (they are discrete analogies of
the Bloch states of the continuum models) is affected by the
nonlinearity in the two cases. One finds that the stationary
solutions of Eq. (\ref{eq:discrete}) at different band edges are
connected by a staggering transformation. It is indeed easy to
prove that if $c_n=\exp(-i\omega t)f_n$ is a stationary solution
of Eq.(\ref{eq:discrete}) for frequency $\omega$ and nonlinearity
$\chi$, then $\tilde{c}_n=\exp(-i(2\omega_0-\omega) t-i\pi n)f_n$
is a solution for frequency $2\omega_0-\omega$ and nonlinearity
$\tilde{\chi}=-\chi$. This means that if at one edge of the band a
mode is  modulationally unstable (and, hence the systems admits
the existence of small-amplitude solitons), at the other edge of
the band for the same nonlinearity the same mode is modulationally
stable and small amplitude solitons cannot exist. For the AL
model, however, one finds  that to a stationary solution $c_n$ for
frequency $\omega$ and nonlinearity $\chi$ corresponds a solution
$\tilde{c}_n$ for frequency $2\omega_0-\omega$ and \textit{the
same nonlinearity} $\chi$. This means that for the AL model it is
possible to have stable small-amplitude solitons at both edges of
the band for a fixed nonlinearity. Actually one can prove that in
this case solitons can exist for all values of the crystal
momentum since the corresponding plane waves are modulationally
unstable in the whole BZ. This means that AL  solitons can slide
along the whole band under the action of a time dependent force
without undergoing any passage through nonexistence region.

While the AL model itself has restricted physical applications, it
shows the route for achieving the stability of gap-solitons in the
whole band with the help of nonlocal nonlinear terms. At the same
time, such terms naturally appear in systems with inhomogeneous
nonlinearities (for a number of nonlocal Hamiltonian lattices and
their properties see e.g.~\cite{ABKKS}). Thus one can expect that
the dynamical localization of continuous gap-solitons could be
achieved in presence of a $x$-dependent nonlinearity. This
conclusion can be reached also from another perspective using the
effective mass picture. As mentioned before, a negative product of
the effective mass and effective nonlinearity is required for all
points of the BZ  to allow the soliton to move along the whole
band (see example in Fig.\ref{fig:ms0}a). While the sign of the
effective mass depends on the band and therefore on the crystal
momentum, the sign of the effective nonlinearity can be changed
only if the nonlinearity is space dependent. For a constant
nonlinearity, indeed,  the soliton will not exist at both edges of
the band, this leading to its spreading and destruction (this is
especially true when the non existence regions are sufficiently
wide with respect to the spectral width of the soliton).\footnote{the
conditions for optimal design of the nonlinear OL to achieve the
conditions of the soliton existence in the whole band was depicted
in Fig.~1a of Ref.~\cite{SKB08}.}
\begin{figure}
\centerline{
\includegraphics{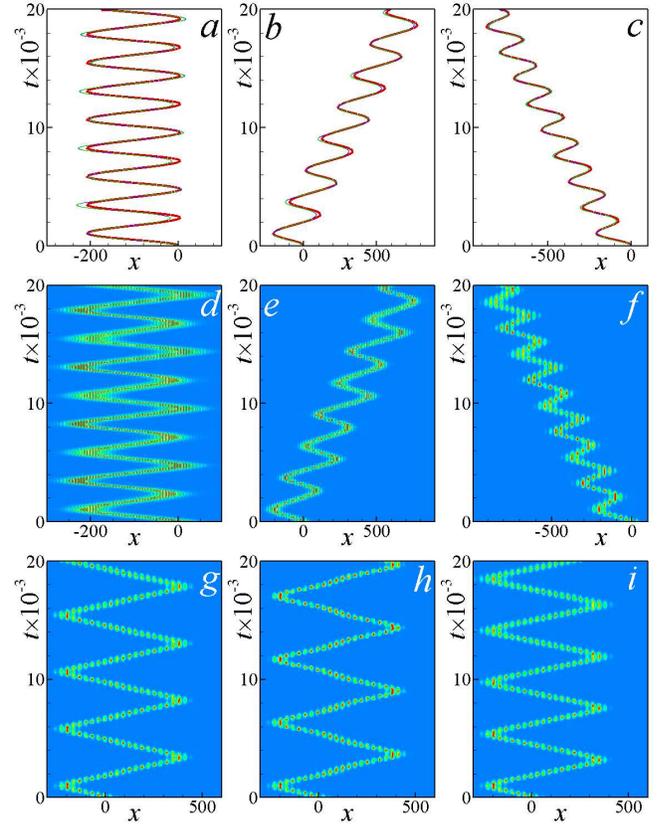}}
\caption{Soliton center position $X$ {\it vs} time $t$ (upper row,
panels a--c) and particle density $|\psi|^2$ {\it vs} time $t$ and
coordinate $x$ (middle and lower rows, panels d--i) for linear and
nonlinear OLs $\cU(x)=-3\cos(2x)$ and $\cG(x)=-0.777+\cos(2x)$,
and external force parameters $\Gamma=-0.001$, $t_0=500.0$ (upper
and middle rows, panels a--f), $t_0=1000.0$ (lower row, panels
g--i), $\Omega\approx 0.001306$ (left column, panels a,d,g),
$\Omega\approx 0.001176$ (middle column, panels b,e,h) and
$\Omega\approx 0.001437$ (lower panel, panels c,f,i). The initial
condition at $t=0$ is a stationary gap-soliton located at $X(0) =
0$ with energy $E=-0.7324$ near the upper edge of the band
($Q(0)=1.0$). In panels a--c bold (red) lines represent
calculation according to the approximation (\ref{eq:X-t2}), and
thin (blue) lines -- direct numerical integration of Eq. (\ref{eq:NLS}).}
\label{fig:X-t}
\end{figure}

The main idea for dynamical localization in the nonlinear case  then is to suppress
(or strongly reduce) non existence regions to allow the
gap-soliton to safely slide along the whole band under the action
of the external force. To show how this can be done we consider
the specific case of the NLS equation with periodic potential and
ac linear force
\begin{eqnarray}
\label{eq:NLS} i\psi_t=-\psi_{xx}+\cU(x)\psi+ \gamma(t)
x\psi +\cG(x)|\psi|^2\psi.
\end{eqnarray}
 In the BEC context the spatial and temporal
variables in Eq.~(\ref{eq:NLS}) are measured in the units of
$d/\pi$ and $\hbar/E_R$, respectively, while the energy is
measured in units of the recoil energy $E_R=\hbar^2\pi^2/(2md^2)$,
where $d$ is the lattice period and $m$ is the atomic mass. In a
typical situations, the external force $\gamma(t)$ arises from the
acceleration of the linear OL. Then Eq. (\ref{eq:NLS}) describes
the condensate in the moving frame. This implies that  the
nonlinear OL should be also accelerated in the same manner as the
linear one. Notice, that Eq.(\ref{eq:NLS}) appears both in  BEC
and in nonlinear optics contexts although the physical contexts
are completely different.

The external force will be assumed to oscillate periodically in
time. While this is not important for the theory, to simplify the
formulas, and to bring the model closer to the experimentally
feasible situations in the theory of matter waves, we consider the
cos-like law $\gamma(t) =\Gamma \cos\left[\Omega(t-t_0)\right]$
with amplitude $\Gamma$ and frequency $\Omega$ (time $t_0$
characterizes the initial phase).

Let us now briefly recall the semiclassical arguments which allow one
to account for the linear force (see e.g.~\cite{Holthaus}).
Identifying the center of a Bloch wave packet, $X$, with the
coordinate of a quasiparticle (which in the nonlinear case is
nothing but a matter wave soliton) one finds that it is described
by the  equations
\begin{eqnarray}
\label{eq:SE} \dot X= v(Q) =\left.\frac {d E(q)}{d q}\right|_{q=Q},
\;\;\;\;\; \dot Q=- \gamma(t),
\end{eqnarray}
where we also introduce the wave-packet center $Q$ in the
reciprocal space, use $v(Q)$ for the Bloch velocity and $E(Q)$ for
the linear OL energy band structure (see Fig.\ref{fig:ms0}a).

Taking into account cos-like dependence of $\gamma(t)$, the
expression for the coordinate   can be represented as
\begin{equation}
X(t)=X_0-2\omega_1\pi\int_{t_0}^t \sin\left\{\pi Q_0-
\frac{\Gamma\pi}{\Omega}\sin[\Omega(\tau-t_0)]\right\}d\tau,
\label{eq:X-t2}
\end{equation}
where we have left only the two leading terms in the Fourier
expansion of the energy $E(q)\approx\omega_0+2\omega_1\cos(\pi
q)$, and use the notations $X(t_0)=X_0$ and $Q(t_0)=Q_0$. In this
case after one period $T=2\pi/\Omega$ of the external force
oscillation, the coordinate of the wavepacket will be
\begin{equation}
X(t_0+T)=X_0 + v_B
\left(Q_0\right)J_0\left(\frac{\Gamma\pi} {\Omega}\right)  T
\label{eq:X-per}
\end{equation}
where $J_0(z)$ is the zero-order Bessel function of the first
kind and $v_B=-2\pi\omega_1\sin(\pi Q_0)$ is the Bloch velocity.
Thus, the mean velocity of the soliton is given by
\begin{equation}
\langle V_S\rangle  = v_B \left(Q_0\right) J_0\left(\frac{\Gamma\pi}
{\Omega}\right), \label{vel}
\end{equation}
i.e. it coincides with the
Bloch velocity at the time $t_0$ of switching on the time
dependent force but scaled by the factor
$J_0\left(\frac{\Gamma\pi} {\Omega}\right)$.

As it can be seen from Eq.(\ref{vel}), if
$\Omega=\Omega_n=\Gamma\pi/z_n$ ($z_n$ is the $n$-th zero of the
Bessel function $J_0(z_n)=0$), then $\langle V_S\rangle =0$ so
that the motion becomes  perfectly periodic  $X(t_0+T)=X(t_0)$ and
the soliton wavepacket is dynamically localized. In Fig.
\ref{fig:ms0}b we depict the soliton mean velocity as a function
of $Q(t_0)$ for fixed parameters of the OLs and fixed frequency
$\Omega$. From this figure it is evident  that the soliton becomes
dynamically localized for any $\Omega$ at the points $Q(t_0)=m$,
$m=-1,0,1$ in the BZ.  The absence of transport at
these points, however, is not due to force resonances but simply
to the Bloch velocity becoming zero (from this point of view the
dynamics at these points should not be considered as dynamical
localization).

To check these predictions, we performed direct numerical
integrations of the NLS equation (\ref{eq:NLS}) to obtain the
evolution of the center of the soliton $X(t)$ when an external
force starts to oscillate at time $t=t_0$
\begin{equation}
\gamma(t)=\left\{\begin{array}{c}\Gamma,\quad t<t_0,
\\ \Gamma\cos\left[\Omega(t-t_0)\right], \quad t\ge
t_0,\end{array}\right.
\end{equation}
The results are depicted in Fig.~\ref{fig:X-t}. As it can be seen
from the panel (a), when $\Omega=\Omega_1$ (corresponds to the
first zero of Bessel function), the dependence, calculated from
approximation (\ref{eq:X-t2}), exhibits the periodic motion.
Nevertheless, the period of this motion is two times less than
that of external force oscillations $2\pi/\Omega_1\approx 4810$,
because the period is imposed mainly by the Bloch oscillations,
whose period is $2/\Gamma=2000.0$. The dependencies in upper
panel, calculated from
direct simulation, exhibit movement, slightly different from
periodic due to the fact, that we restricted expansion of $E(q)$
into Fourier series by zeroth and first harmonics only. In cases,
when $\Omega$ differs from $\Omega_n$ (in panel (b)
$\Omega=0.9\Omega_1$ and in panel (c) $\Omega=1.1\Omega_1$), the
soliton movement looses its periodicity.

In panels d--i of Fig. \ref{fig:X-t} we depict the time evolution
of the matter density as obtained from direct integration of the
NLS equation. The panels d,e,f correspond to panels a,b,c, respectively.
Notice that in all the cases the soliton is long lived as a
consequence of a proper design of the OLs. In the nonresonant case
$J_0\left(\frac{\Gamma\pi} {\Omega}\right) \ne 0$ (panels e,f) the
mean velocity of the soliton is $\langle V_S \rangle  \approx
0.03722$ (panel e) and $\langle V_S\rangle \approx -0.04192$, this
agreeing very well with the value predicted by Eq. (\ref{vel})
(see Fig. \ref{fig:ms0}).

In panels g--i of Fig. \ref{fig:X-t} we have shown that
localization can also be induced in nonresonant case if the
initial Bloch velocity at time $t_0$ is zero, in agreement with
what predicted by Eq. (\ref{vel}) (notice that $t_0=1000$ and the
external force oscillation starts when the soliton is at the
bottom of the band $Q_0=0$). In this case the dynamical
localization can occur for any frequency $\Omega$ of the external force,
as expected from Eq.(\ref{eq:X-per}).
\begin{figure}
\centerline{
\includegraphics{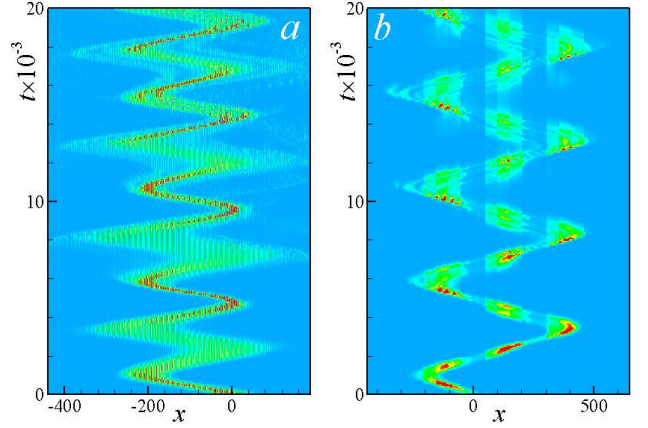}}
\caption{Particle density $|\psi|^2$ {\it vs} time $t$ and
coordinate $x$  for linear OL $\cU(x)=-3\cos(2x)$, constant
nonlinearity $\cG(x)=1$, and external force parameters
$\Gamma=-0.001$, $\Omega\approx 0.001306$, $t_0=500.0$ (panel a),
$t_0=1000.0$ (panel b). The initial condition at $t=0$ is a
stationary gap-soliton located at $X(0) = 0$ with energy
$E=-0.7324$ near the upper edge of the band ($Q(0)=1.0$). }
\label{fig:const-nonl}
\end{figure}

 To demonstrate the importance of the nonlinear OL
for the stability of dynamical localization of solitons, we
display in Fig.\ref{fig:const-nonl}a,b, the
dynamics of soliton under the periodic external force with the
same parameters, as those of Fig.\ref{fig:X-t}d,e, except for the
case of coordinate-independent nonlinearity $\cG(x)=1$. As evident
from Fig.\ref{fig:const-nonl}, in the case of constant
nonlinearity solitonic oscillations in real space are not stable.
This happens due the above-described mechanism of soliton
distortion in the modulationally stable regions of BZ. This
situation differs considerably from the case, depicted in
Fig.\ref{fig:X-t}, where special shape of nonlinear OL results in
narrow width of the modulationally stable regions in the BZ (see
Ref.\cite{SKB08}) so that they do not affect considerably on the
soliton stability during its propagation in the real and
reciprocal spaces.

\begin{figure}
\centerline{
\includegraphics{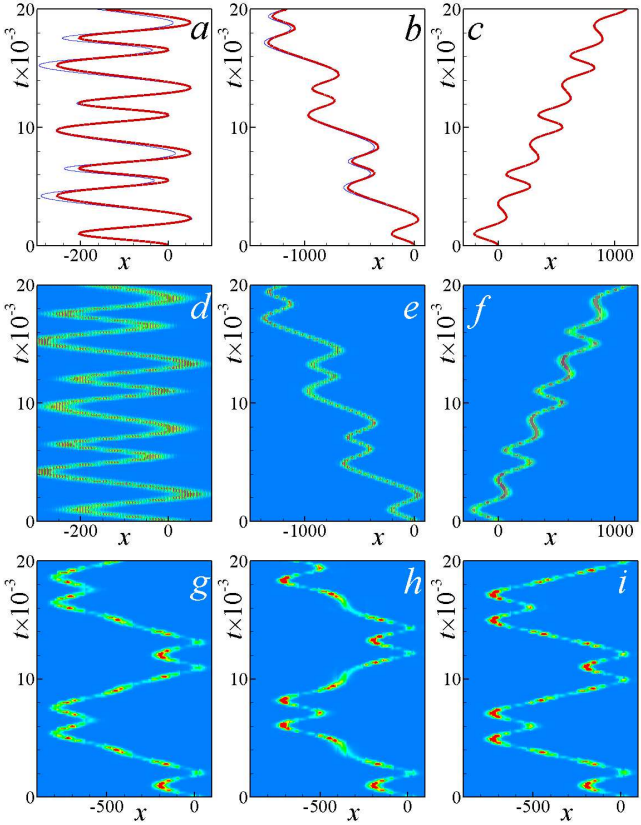}
} \caption{The same as Fig.\ref{fig:X-t}, but for $\Omega\approx 0.000569$
(left column, panels a,d,g), $\Omega\approx 0.000512$
(middle column, panels b,e,h) and $\Omega\approx 0.000626$ (lower
panel, panels c,f,i).} \label{fig:Xsr-t}
\end{figure}

The phenomenon of dynamical localization can occur also, when
external force oscillation frequency corresponds to the higher
roots of Bessel function.  In other words, within
the framework of the meanfield theory, there exist no lower limit
on the frequency leading to the dynamical localization. Using the
well known asymptotic of the Bessel functions one obtains the
behavior of the low frequency limit (recall that they are measured
in the units $E_r/\hbar$): $\Omega\sim\Gamma/\left(n-\frac
14\right)$.  Reducing the frequency however, generally speaking
leads to larger amplitudes of soliton oscillations and more
sophisticated trajectories. An example of this is reported in
Fig.\ref{fig:Xsr-t} where the dynamical localization is shown for
the case $\Omega=\Omega_2$. In this case, however, we see a small
discrepancy between a tight-binding approximation and direct
numerical simulation of the NLS system  in Eq.(\ref{eq:NLS})
(Fig.\ref{fig:Xsr-t}a).

 In closing this letter we wish to remark that
periodically driven OLs have also been recently used
experimentally to achieve coherent control of matter
waves~\cite{Zenesini} and theoretically to investigate phase
transition from superfluid to a Mott insulator by means of
Bose-Hubbard model~\cite{EWH05}. In both cases an effective
renormalization of the tunneling matrix element in the
Bose-Hubbard hamiltonian was assumed. In the present  Letter we
have dealt with a quite different situation in which  each lattice
site is populated by many atoms so that the mean field approach is
perfectly valid. The presence of an  inhomogeneous
nonlinearity in our case permits the existence of gap soliton for
all points in the BZ. The gap-soliton dynamics is then well
described by the semiclassical equation of motion and is very
stable, without any visible increase of the local density, this
ruling out any possibility for occurrence of phase transitions or
failure of the 1D approximation (the latter occurring when the
two-body interaction energy become comparable to the transverse
kinetic energy).

In conclusion, we have shown for the first time the possibility of
{\em nonlinear} dynamical localization of matter wave solitons  by
means of a spatial modulation of the nonlinearity. In a BEC this
can be achieved by changing the scattering length by means of
optically (or magnetically) induced Feshbach resonances while the
ac force can be generated by a time periodically accelerated OL.
For other possible applications of the theory, which are arrays of
coupled nonlinear waveguides, the spatial change of the
nonlinearity can be made by changing the refractive index in the
fibers along the array with dopants, while the periodic linear
force could be applied by curving the fibers along the propagation
direction.

The phenomenon of dynamical localization of solitons in OL is
expected then to be observed both in BEC and in nonlinear optical
systems.

\smallskip
Y.V.B. acknowledges support from FCT, Grant
No. SFRH/PD/20292/2004.   M.S. wish to
acknowledge the hospitality received at the Centro de F\'{\i}sica
Te\'orica e Computacional, Universidade de Lisboa, and partial
support from a MIUR-PRIN-2008 initiative.
The cooperative work was partially supported by the bilateral
agreement FCT(Portugal)/CNR(Italy).

\end{document}